\documentstyle[preprint,aps,eqsecnum]{revtex}
\tighten
\begin{document}
\preprint{PAR-LPTHE/96-24, EHU-FT/9601}
\draft
\title{Planetoid string solutions in  $ 3 + 1 $
axisymmetric  spacetimes}
\author{H.J. de Vega\cite{emLPTHE}}
\address{Laboratoire de Physique Th\'eorique et Hautes \'Energies
\\
Universit\'es Paris VI-VII - Laboratoire associ\'e au CNRS n$^{\rm o}$ 280
\\
Tour 16, 1er. \'et., 4, Place Jussieu, 75252 Paris Cedex 05
\\ FRANCE}
\author{I.L.
Egusquiza\cite{emUPV}}
\address{Fisika Teorikoaren Saila\\
Euskal Herriko Unibertsitatea\\
 644 P.K. - 48080 BILBAO\\
SPAIN}
\date{\today}
\maketitle
\begin{abstract}
The string propagation equations in axisymmetric spacetimes are
exactly solved by quadratures for a planetoid Ansatz. This is a  straight
non-oscillating string, radially disposed, which rotates uniformly around the
symmetry axis of the spacetime. In Schwarzschild
black holes, the string stays outside the
horizon pointing towards the origin. In de Sitter spacetime the
planetoid rotates around its center.
We quantize semiclassically these solutions and analyze the
 spin/(mass$^2$) (Regge) relation for the planetoids, which turns out to be
non-linear.
\end{abstract}
\pacs{11.25.-w, 04.70.-s, 98.80.Cq }

\section{Introduction and Motivations}

The systematic investigation of strings   in curved spacetimes
started in \cite{plb} has uncovered a variety of new physical phenomena
(see \cite{erice,nueview} for a general review). These results  are
relevant both for
fundamental (quantum) strings and for cosmic strings, which behave in
an  essentially classical  way.

The study of classical and semiclassical strings in curved backgrounds will
provide and is indeed providing us with a better comprehension of what a
consistent string theory and gravity theory entail. In this
context we place
the present paper, which continues the line of research set by
\cite{plb}.

Among the heretofore  existing analysis of the motion of classical strings in
gravitational backgrounds a special place is to be granted to exact solutions,
usually obtained by means of separable ans\"atze
(non-separable exact solutions were systematically constructed  for de
Sitter spacetime \cite{cdms}).
Such are the circular string
ansatz \cite{letelier,hjile}, which for stationary axially symmetric spacetimes
reduces the nonlinear equations of string motion to an equivalent
one-dimensional dynamical system
\cite{schwads}, or the stationary string ansatz \cite{demirm953}.

In this paper we examine a different ansatz, which we have called the planetoid
ansatz, in  stationary axisymmetric $ 3 + 1 $ spacetime backgrounds.
The planetoid solutions are straight  non-oscillating string solutions
that rotate uniformly around the symmetry axis of the spacetime.
In Schwarzschild black holes,
they are  permanently pointing towards $ r = 0 $ while
they rotate outside the horizon. In de Sitter spacetime the planetoid
rotates around its center.

We call our ansatz planetoid since it generalizes to strings the
bounded circular orbits of point particles in such spacetimes. In the
case of the
Schwarzschild geometry the planetoid string solutions presented here
are generalization of the circular orbits of planets.

We will show how our planetoid
ansatz produces either orbiting strings with bounded world-sheet and
length or  strings of unbounded length.
The main competing physical forces in the context of this ansatz are the
attraction of gravity, the centrifugal force, and the string tension. The
combination of these three causes in different proportions  produce
different effects, as we will now see.

It should be noted that the effects mentioned  take place even when the
 gravitational field acting on the
string is not strong. They   are due to the non
local character of the string.

We quantize semiclassically the planetoid string solutions using the
WKB method  adapted to periodic string solutions \cite{dvls}.
We obtain in this way their  masses as a function of the angular momentum. Such
relations are non-linear and can be considered as a (generalized)
Regge trajectory [See figs. 1 and 2].

\section{Equatorial planetoid ansatz}

\subsection{The ansatz and the string equations of motion}

We consider our classical strings propagating in $ 3 + 1 $ dimensional
 stationary axisymmetric spacetime. For simplicity, we restrict in
 this paper to strings propagating in  the
equatorial plane $ z = 0 $.  We can thus restrict ourselves to the
$2+1$ metric with  line element of the form
\begin{equation}
{\rm d}s^2=g_{tt}(r)\;  {\rm d}t^2 +g_{rr}(r)\;{\rm d}r^2 + 2 \;
g_{t\phi}(r)\; {\rm  d}t\, {\rm d}\phi + g_{\phi\phi}(r)\; {\rm d}\phi^2\,.
\end{equation}

Let $\tau$ and $\sigma$ be the time-like and
space-like world-sheet coordinate respectively in the conformal
gauge. Under the ansatz
\begin{eqnarray}
t & = & t_0 + \alpha\, \tau\,,\nonumber\\
\phi & = & \phi_0 + \beta\,\tau\,,\nonumber\\
r & = & r(\sigma)\,,\label{ansatz}
\end{eqnarray}
the equations of motion for a string in this background are given by the
following one-dimensional equivalent system:
\begin{eqnarray}
\left({{{\rm d}r}\over{{\rm d}\sigma}}\right)^2 + g^{rr}\big[\alpha^2
\; g_{tt}
+2 \, \alpha \,\beta  \, g_{t\phi} & + &
\beta^2 \, g_{\phi\phi}\big]=\nonumber\\
\left({{{\rm d}r}\over{{\rm d}\sigma}}\right)^2
+V(r) & = & 0\,.\label{laecuacion}
\end{eqnarray}
The function
$r(\sigma)$  will then be given by the zero energy motion in
$\sigma$ ``time'' of $r$ under the potential $V(r)=\alpha^2  \,
g^{rr}\left[g_{tt}
+ 2 \,\lambda  \,g_{t\phi}+
\lambda^2 \, g_{\phi\phi}\right]$, with $\lambda=\beta/\alpha$.

Quite obviously,
the movement of the string will be periodic. The physical period
$T$ in coordinate time $t$  relates to
$\lambda$ through
$$
T={{2\pi}\over {\lambda}} \; .
$$
It  will prove useful to introduce the `physical' potential
$$
\tilde V(r)=V(r)/\alpha^2=g^{rr}\left[g_{tt}
+ {{4\pi}\over T }\,  g_{t\phi}+
 {{4\pi^2}\over T^2}\; g_{\phi\phi} \right] \; ,
$$
since it only depends on the physical parameter $T$.

The boundary conditions for open strings, namely,
$\partial X_\mu/\partial\sigma=0$ at the ends of the string, are naturally
fulfilled by this ansatz.

Following
\cite{chgordon}, we see that this is the only ansatz that separates
variables,
lets strings be dynamical, and respects the open string
boundary conditions, when $r=r(\sigma)$ is chosen.

Note also that this ansatz differs from the circular string ansatz (i.e.,
$t=t(\tau)$,
$\phi=\phi_0 +\nu\sigma$, $r=r(\tau)$) in the dependence of $r$ in the
space-like world-sheet (conformal) coordinate and in the form of the equivalent
one-dimensional energy equation, which for this later case reads $\dot r^2+
g^{rr}\left[\mu^2 g^{tt} +\nu^2 g_{\phi\phi}\right]$, where the dot stands for
the derivative with respect to $\tau$.

The invariant size of the planetoid string is given by the substitution of the
ansatz in the line element:
\begin{equation}\label{talla}
{\rm d}s^2=g_{rr}\left({{{\rm d}r}\over{{\rm d}\sigma}}\right)^2\left(-{\rm
d}\tau^2+{\rm d}\sigma^2\right)\,.
\end{equation}

\subsection{Energy and angular momentum}
It is well known that the definition of a stress-energy tensor for an extended
object in general relativity is no mean task \cite{dixon}. In the case at hand,
however, there exists a favored time coordinate, for which a Killing vector
exists ($\partial/\partial t$). This allows us to define clearly what is meant
as energy,  \cite{dixon}: ${\cal E}=\alpha/\alpha'$.  

Similarly, the existence of the Killing
vector $\partial/\partial\phi$, associated with the rotational symmetry, allows
for the definition of an angular momentum about the axis. In particular, this
is performed as follows: the function $\phi(\sigma,\tau)$ appears in the string
Lagrangian only through its derivatives, whence the conserved
world-sheet current is obtained by Noether's theorem
$$
J_{\mu} = {2 \over {\pi \alpha'}}\; \left[ g_{t\phi} \; \partial_{\mu}t
+ g_{\phi\phi}  \; \partial_{\mu}\phi \right] \; .
$$
The integration of this current provides us with the string angular
momentum $J$, 
$$
J \equiv \int J_{\tau} \; d\sigma =  {2 \over {\pi \alpha'}}\;
\int_{{r_{\rm min}}}^{{r_{\rm max}}}{\rm d}r\,
{{g_{t\phi} + {{2\pi}\over T}\;  g_{\phi\phi}}\over{\sqrt{-\tilde
V(r)}}}\, .
$$
where we used eq.(\ref{laecuacion}) and $r_{\rm min}$ and $r_{\rm
max}$ denote the minimum and maximum radius reached by the string,
respectively. 

\subsection{General expressions and quantization condition}

We will collect here the expressions for the physical string magnitudes:
angular momentum $J$, classical
action for solutions $S_{\rm cl}$, mass $ m $, and reduced action $
W(m) $. The mass will be defined as
$m:=-{\rm d}S_{\rm cl}/{\rm d}T$, with $T$ the period. The reduced action
\cite{dvls} is thus obtained as $W(m)=mT(m) +S_{\rm
cl}\left(T(m) \right)$.
The quantization condition will read $W(m)=2\pi n$ (in units with $\hbar
=1$).

For the case at hand, closed expressions in terms of quadratures can be
obtained for all these quantities, as follows:
\begin{eqnarray}\label{expresiones}
S_{\rm cl}(T) & = & -{{2T}\over{\pi\alpha'}} \int_{{r_{\rm min}}}^{{r_{\rm
max}}}{\rm d}r\,g_{rr}\sqrt{-\tilde V(r)}\,,\cr\cr
W & = & {{4}\over{T\alpha'}}\int_{{r_{\rm min}}}^{{r_{\rm
max}}}{\rm d}r\,{{Tg_{t\phi} + 2\pi g_{\phi\phi}}\over{\sqrt{-\tilde
V(r)}}}\,,\cr\cr
m & = & {{W-S_{\rm cl}}\over{T}}\,,\cr\cr
J & = & {{W}\over{2\pi}}\,.
\end{eqnarray}

As is immediately obvious from these
expressions, it is not necessary to have the solution $r(\sigma)$ in a closed
form for the quantities indicated to be evaluated, and in what follows we will
not use the explicit expressions for  $r=r(\sigma)$,
which, after all, is dependent on the parametrization of the world-sheet. It
should be noted that the previously mentioned quantization condition
[$ W(m)=2\pi n $] is equivalent for this class of solutions to
$J=n$. This should be interpreted as a consistency check of the
semiclassical quantization being performed.

The invariant string length at a fixed time $ t $ follows from
eq.(\ref{talla})
\begin{equation}\label{loninv}
s = \int_{{r_{\rm min}}}^{{r_{\rm max}}}{\rm d}r\,\sqrt{g_{rr}} \; .
\end{equation}

\section{Explicit solutions and their analysis}
\subsection{Minkowski spacetime}
In order to improve our understanding of  the physical meaning of the
solutions being examined, let us take the easy Minkowski case, for which
$g_{tt}=-1$,
$g_{rr}=1$,
$g_{t\phi}=0$, and
$g_{\phi\phi}=r^2$. Equation (\ref{laecuacion})  then becomes
$$
\left({{dr}\over {d\sigma}}\right)^2+ \lambda^2 r^2-1=0\,,
$$

The solution is immediate:
$$
r={T \over {2\pi}}\; \left|\cos\left({{2\pi}\over T} \sigma \right)\right| \; ,
$$
where $ 0 \leq \sigma \leq T/2 $.
In cartesian coordinates,
$$
x = {T \over {2\pi}}\; \cos\left({{2\pi\tau}\over T} \right)\, \cos
\left({{2\pi\sigma}\over T} \right)\,
$$
$$
y  = {T \over {2\pi}}\; \sin\left({{2\pi\tau}\over T} \right)\, \cos
\left({{2\pi\sigma}\over T} \right)\, .
$$
It is easy to see that this is  a string of length $ T/\pi $
rotating around its middle point which coincides with the origin of
coordinates.

The action, reduced action, mass and angular momentum are therefore
\begin{eqnarray}
S_{\rm cl}(T) & = & -W=-{{T^2}\over{2\pi\alpha'}}\,,\cr \cr
m & = & {T \over {\pi\alpha'}} \, ,\cr \cr
J & = & {{T^2}\over {4\pi^2\alpha'}}\,,
\end{eqnarray}
from which the relation follows
\begin{equation}
\alpha' m^2 = 4J\,.
\end{equation}
It should be noted that this relation differs from the standard one by a factor
4. This is due to the different normalization of the string tension parameter
$\alpha'$.

\subsection{Static Robertson-Walker spacetimes}

As a first curved spacetime, we  examine the static Robertson-Walker
universe, with line element
\begin{equation}
{\rm d}s^2= -{\rm d}t^2 + {{{\rm d}r^2}\over{1-\kappa r^2}} + r^2 {\rm
d}\phi^2\,.
\end{equation}
If $\kappa<0$, the potential
$$
{\tilde V} = {{\left({{2\pi r}\over T}\right)^2 - 1 } \over {1 -\kappa r^2}}
$$
is smaller than zero if $0<r<T/2\pi$, as in
Minkowski spacetime. On the other hand, were we to take $\kappa>0$, the
number of
possible types of solutions increases. Consider first $\kappa<0$. Let
$\nu=T\sqrt{-\kappa}/2\pi$, and $\mu=\nu/\sqrt{\nu^2+1}$.

Our computations result in
\begin{eqnarray}
S_{\rm cl}(T) & = &
-{{4T}\over{\pi\alpha'\sqrt{-\kappa}}}\; {1\over{\mu}}\; \left[K(\mu)
-E(\mu)\right]\,,\cr \cr
W & = & {{4T}\over{\pi\alpha'\sqrt{-\kappa}}}\; \left[{1\over{\mu}}\; E(\mu) +
{{\mu^2-1}\over{\mu}}K(\mu)\right]\,,\cr \cr
m & = & {{4}\over{\pi\alpha'\sqrt{-\kappa}}}\;  \mu \, K(\mu)\,,
\end{eqnarray}
where $K$ and $E$ are complete elliptic integrals of the first and second kind
respectively, with the elliptic modulus as their argument.

Let us now pass to the $\kappa>0$ situation. There are two classes of
solutions: those that extend from $0$ to ${\rm min}(r_T,r_\kappa)$, and those
from  ${\rm max}(r_T,r_\kappa)$ to infinity, where $r_T=T/2\pi$ and
$r_\kappa=1/\sqrt{\kappa}$. The second class of solutions lead to infinite
reduced action. As to the first class, computations yield
\begin{eqnarray}
S_{\rm cl}(T) & = & -{{8r_\kappa^2}\over{\alpha'}}\left[
E\left({{r_T}\over{r_\kappa}}\right) +
\left({{r_T^2}\over{r_\kappa^2}}-1\right)
K\left({{r_T}\over{r_\kappa}}\right)\right]\,,\cr \cr
W & = &
{{8r_\kappa^2}\over{\alpha'}}\left[K\left({{r_T}\over{r_\kappa}}\right)
-E\left({{r_T}\over{r_\kappa}}\right)\right]\,,\cr \cr
m & = & {{4 r_T}\over{\pi\alpha'}}\; K\left({{r_T}\over{r_\kappa}}\right)\,,
\end{eqnarray}
for the case $r_T<r_\kappa$, and
\begin{eqnarray}
S_{\rm cl}(T) & = & -{{8r_\kappa r_T}\over{\alpha'}}
E\left({{r_\kappa}\over{r_T}}\right)\,,\cr \cr
W & = & {{8r_\kappa
r_T}\over{\alpha'}}\; \left[K\!\left({{r_\kappa}\over{r_T}}\right)
-E\left({{r_\kappa}\over{r_T}}\right)\right]\,,\cr \cr
m & = & {{4
r_\kappa}\over{\pi\alpha'}}\; K\!\left({{r_\kappa}\over{r_T}}\right)\,,
\end{eqnarray}
for the case $r_T>r_\kappa$.

We see that the string angular momentum $ J = W/(2\pi) $ {\bf is not}
proportional to $ m^2 $ yielding a non-linear Regge trajectory.

In the  $\kappa\to 0^+$ limit,  we have
\begin{eqnarray}
W & \buildrel{ \kappa \to 0  }\over =  & {{T^2}\over{2\pi\alpha'}}\left(1+
{{3 T^2}\over{32\pi^2}}\; \kappa + \ldots\right)\,,\nonumber\\
m & \buildrel{ \kappa \to 0  }\over =  & {{T}\over{\pi\alpha'}}\left(1+
{{T^2}\over{16\pi^2}}\;\kappa + \ldots\right)\,,\label{aproximado}
\end{eqnarray}
and, consequently,
$$\alpha' m^2 \buildrel{ \kappa \to 0  }\over = 4n \left(1+ {{n \alpha'}\over8}
\; \kappa +\ldots\right)\,.$$
In the  $\kappa\to 0^+$ limit we find a linear Regge trajectory,
recovering  the previous results for Minkowski spacetime.

\subsection{Cosmological and black hole spacetimes}
Let us consider spacetimes with the generic form $g_{rr}=1/a(r)=-1/g_{tt}$,
$g_{t\phi}=0$. The potential $\tilde V$ is then given
by
$\tilde V(r)=a(r)\left(\lambda^2 g_{\phi\phi}-a(r)\right)$. Since the
``motion''  of
$r$ in $\sigma$ can only take place when $\tilde V(r)<0$,  we have to
determine the 
zeroes of $a(r)$ and of $\lambda^2 g_{\phi\phi}-a(r)$, together with the
asymptotics in the different physical regions.

\subsubsection{de Sitter spacetime}
Included within this set of metrics we find the de
Sitter metric, for which $a(r)=1-H^2r^2$ and $g_{\phi\phi}=r^2$. The radius of
the horizon, $r_H$, is given by $ r_H=1/H $. Thus,

$$
{\tilde V}(r) = (1-H^2r^2)\left\{ \left[ H^2 + \left( {{2\pi}\over T}
\right)^2 \right] \; r^2 - 1 \right\} \; .
$$

The zeroes of the potential $V$ in this case are
$r_H$ and $r_H/\sqrt{1+\left( {{2\pi}\over {HT}}\right)^2}$.

There are two types of planetoid
strings: those of infinite length that are to be found outside the horizon,
and those completely within the horizon, that are of finite length. Let us
concentrate on the later. The maximum radius is
$ r_{\rm max} = r_H/\sqrt{1+(\lambda/H)^2} $ and $ r_{\rm min} = 0 $.
This is a string rotating around its middle point located precisely
at $ r = 0 $.

The integrals to be performed are complete elliptic integrals, with elliptic
modulus
$$
k=HT/\sqrt{H^2T^2 + 4\pi^2} \; .
$$
Let $ k'=\sqrt{1-k^2} $. Our computations result in the following:
\begin{eqnarray}\label{dS}
S_{\rm cl}(T) & = & -{{8 }\over{k'\alpha'\, H^2}}\left[E(k) -
{k'}^2K(k)\right]\,,\cr \cr
W & = &  2 \pi \, J ={{8 }\over{\alpha'\, H^2}}\;k'\, \left[K(k) -
E(k)\right]\,,\cr \cr
m & = & {4\over{\pi\, H \, \alpha'}}\,  k \,E(k)\,.
\end{eqnarray}
It is  here obvious that $ J $ is not proportional to $ m^2 $.

For small $HT$, the quantization condition reads $T^2\sim
4\pi^2 n\alpha'$, as in flat spacetime, and the mass of the string is in this
case (compare with \cite{dvls})
\begin{equation}
\alpha'm^2\simeq 4n -7\, H^2 \alpha' n^2 +\cdots \; .
\end{equation}

It follows from eqs.(\ref{dS}) that $ k $ is a two-valued function of
$ W $ and hence of $ n $. Therefore, there are {\bf two } values of $
m $ for each $ n $. This is easy to see from the behaviour of $ W $
for $ k \to 0 $ and for $ k \to 1$. $ W $ vanishes in both cases.
$$
W  \buildrel{ k \to 1  }\over = { 8 \over {  \alpha'\, H^2}} \; k' \;
\left( \log{4 \over {k'}} - 1 \right) + O(k'^3 \, \log k' ) \; ,
$$
$$
W  \buildrel{ k \to 0  }\over = {{ 2\pi } \over {  \alpha'\, H^2}} \;
k^2 + O(k^4) \; .
$$
 There is  a maximum on the values $n$ can take, given by
\begin{equation}
n\leq n_{max} \equiv 0.616\,{1\over{\alpha'\, H^2}} \,.
\end{equation}
This $  n_{max} $ correspond to the maximal planetoid mass.

The first branch yields  masses in the range
$$
0 \leq m \leq m_{\rm max}=1.343\ldots\; {1 \over { H \, \alpha'}} \;,
$$
and the second branch in the range
$$
{4\over {\pi\, \alpha'\, H}} \leq m \leq m_{\rm max}=1.343\ldots\; {1
\over { H \, \alpha'}} \; .
$$
[Notice that $ \frac4{\pi} = 1.2733\ldots $].

We find from eq.(\ref{loninv}) for the invariant string length
$$
s = \frac2{H}\; {\rm arcsin}{1 \over {\sqrt{1 +
\left({{2\pi}\over{HT}}\right)^2 }}}\; .
$$
$s$ takes its maximum value $ {{\pi}\over {H}} $ for the lightest
states in the second branch $ k \to 1 , \; m \to {4\over {\pi\, \alpha'\,
H}}$. The shorter planetoids $ s \simeq {T \over {\pi}} , \; k \to
0 $ correspond to the lightest states in the first branch.

With  respect to the infinite length planetoid solutions (that is to say, those
restricted to be outside the horizon), the corresponding action, reduced action
and mass are all infinite.

\subsubsection{Anti-de Sitter spacetime}

In this case, $a(r)=1+H^2r^2$ and $g_{\phi\phi}=r^2$. Only for a
 restricted set
of values of $\lambda$ will there be a change of sign in $V$, since only if
$\lambda^2>H^2$ will there be a zero of $V(r)$, namely at
$1/\sqrt{\lambda^2-H^2}$. Therefore, lower values of $\lambda$ correspond to
strings of infinite length, whereas those strings for which $\lambda^2>H^2$
will be of finite length. They will rotate  around its middle point
located precisely at $ r = 0 $ with period $ T \; , \; 0 < T <
{{2\pi}\over H} $.

The results for this spacetime are as follows, where
$k=HT/(2\pi)=H/\lambda$:
\begin{eqnarray}
S_{\rm cl}(T) & = & -{{8 }\over{H^2\, \alpha'}}\left[K(k) -
E(k)\right]\,,\cr \cr
W & = & {{8}\over{(H\, k')^2\alpha'}}\left[E(k) - (k')^2
K(k)\right]\,, \cr \cr
m & = & {{4 }\over{\pi\, H \, \alpha'}}{{k}\over{(k')^2}} \, E(k)\,.
\end{eqnarray}

 In this case $W$ is a monotonous function of $T$, and so is $m$, so the
doubling of mass eigenvalues found in de Sitter spacetime is not present
here.

For the low-lying mass states we find
$$
\alpha'm^2\simeq 4n -H^2 \alpha' n^2 +\cdots \; .
$$
There is no upper bound in the mass spectrum for anti-de Sitter
spacetime. For large masses we find
$$
m \simeq2 n H \; , \; n>>1 \; .
$$
The heavy states spacing  is given by $ H $ whereas the small mass
spacing is determined by $ ( \alpha' )^{-\frac12} $.

\subsubsection{Schwarzschild black hole}

For the Schwarzschild black hole $a(r)=1-2M/r $ and $g_{\phi\phi}=r^2
$, where $ 2 M $ stands for the Schwarzschild radius.

There will be positive zeroes of $V$ other than  that at
$2M$ if and only if $16 \pi^2 M^2/T^2\leq4/27$. Of the two additional zeroes
in
this case, one will be placed between $1$ and $1.5$, and the other will be
larger
than $1.5$ in units of $ 2 M $. In the extreme case  $16 \pi^2
M^2/T^2=4/27$ the two will coalesce onto
$r_0=1.5$, which is the minimal (unstable) radius for a circular null
geodesic \cite{chandra}. As we turn $T$ to larger values, one of
the zeroes runs to $1$, and the other out to infinity, these extreme values
being reached for
$2\pi/T=0$, thus corresponding to an infinite static string from the horizon
to infinity \cite{fszh}.

Let us choose the following parametrization for $T$, and consequently for the
roots $r_i$ of $V$, with $r_i=2Mx_i$:
\begin{eqnarray}\label{raices}
T & = & {{6\pi M\sqrt{3}}\over{\cos(3s)}}\,,\\ \cr
x_1 & = & -{{3\cos s}\over{\cos(3s)}}\,,\cr\cr
x_2 & = & {{3}\over{2\cos(3s)}}\left(\cos s -\sqrt{3}\sin s\right)\,,\cr\cr
x_3 & = & {{3}\over{2\cos(3s)}}\left(\cos s +\sqrt{3}\sin
s\right)\,,\nonumber
\end{eqnarray}
whence $(r=2M x)$
\begin{eqnarray}
\tilde V(r) & = & {1\over{x^2}}(x-1)\left( {{4\cos^2(3s) x^3}\over{27}} -
 x + 1\right)\cr
& = & {{4\cos^2(3s)}\over{27x^2}}(x-1)(x-x_3)(x-x_2)(x-x_1) \,.
\end{eqnarray}
The parameter $ s $ is a function of $ T/M $ as defined by
eq.(\ref{raices}). $ s $ runs from $0$ to $\pi/6$, and the roots are ordered as
$x_1<1<x_2<x_3$. The planetoid string extends from $ r = 2 m x_2 $ to
$   r = 2 m x_3 $. Its invariant length follows from eq.(\ref{talla})
$$
s = 2M [ f(x_3) -  f(x_2)] \; ,
$$
where
$$
f(x) = \sqrt{x(x-1)} -{\rm ArgTh}\sqrt{1 - \frac1{x}}\;.
$$

The classical, reduced action and mass are then integrals expressible in
terms of elliptic integrals of modulus
$$
k^2={{(x_3-x_2)(1-x_1)}\over{(x_3-1)(x_2-x_1)}} \; .
$$
 The
explicit expressions are not by themselves very illuminating, since they
involve combinations of elliptic integrals of different kinds; as a
simple exponent, we have
\begin{eqnarray}
m & =
&{{T}\over{\pi^2\alpha'}}{{2(x_2-1)}\over{\sqrt{(x_3-1)(x_2-x_1)}}}\times\cr
& & \quad\Pi\left({{x_3-x_2}\over{x_3-1}},
\sqrt{{{(x_3-x_2)(1-x_1)}\over{(x_3-1)(x_2-x_1)}}}\right)\,.
\end{eqnarray}
We use, as before, the notation of ref.\cite{gr}.

An important point is that there is a minimum value for the reduced
action and for the mass, corresponding to
$T_{\rm min}=6\pi\sqrt{3} M$, as follows:
\begin{eqnarray}
W_{\rm min} & = & {{36\pi M^2}\over{\alpha'}}\,,\cr \cr
m_{\rm min} & = & {{2\sqrt{3} M}\over{\alpha'}}\,.
\end{eqnarray}
The classical action for this configuration vanishes.

The period $ T $ has no upper bound. For large $ T $ we find
very long strings with
$$
W \buildrel{ T \to \infty  }\over =  { 1 \over {6 \pi^3 \, \alpha'
M}}\; T^3 \quad , \quad
m \buildrel{ T \to \infty  }\over =  { 1 \over {4 \pi^3 \, \alpha' M}}\; T^2
 \quad , \quad s  \buildrel{ T \to \infty  }\over = { T \over {2 \pi}}
- M \log{T \over {\pi M}} + O(1) 
$$
and the mass spectrum
$$
 (\alpha' M)^{1/3} m \buildrel{ n \to \infty  }\over =
 \left(\frac9{4\pi}\right)^{1/3}\; n^{2/3} \; .
$$
The Regge trajectory $W(m)$ is  well behaved, and we portray it in
Fig. 1.
\subsubsection{Schwarzschild black hole in de Sitter spacetime}
We shall now find competing effects due to the presence of one cosmological
and one black hole horizons. The function $ a(r) $ equals $1-2M/r -
H^2 r^2$, and $g_{\phi\phi}=r^2$. We are presented with three cases:
\begin{itemize}
\item{}
$1/(27 M^2)>(4\pi^2/T^2+H^2)$ (and, a fortiori, $1/(27 M^2)>H^2$); the positive
roots of the potential are the cosmological horizon, the black hole
horizon, and two others, which we examine later.
\item{} $(4\pi^2/T^2+H^2)>1/(27 M^2)>H^2$, when only strings inside
the black hole horizon and outside the cosmological horizon are present within
our ansatz.
\item{} $H^2>1/(27 M^2)$, which entails that there is no horizon and no
strings of the form of our ansatz.
\end{itemize}
We shall now study the first of these cases, when there are four positive
roots of the potential $V$, using a parametrization analogous to the one
before. Let
\begin{eqnarray}
x & = & r/(2M)\,,\cr \cr
H & = & {{\cos(3s_1)}\over{3\sqrt{3}M}}\,,\cr \cr
T & = & {{6\sqrt{3}\pi M}\over{\sqrt{\cos^2(3s_2)-\cos^2(3s_1)}}}\,,\cr \cr
x_{\rm neg} & = & -{{3\cos(s_1)}\over{\cos(3 s_1)}}\,,\cr \cr
x_S & = & {{3}\over{2\cos(3
s_1)}}\left(\cos(s_1)-\sqrt{3}\sin(s_1)\right)\,,\cr \cr
x_H & = & {{3}\over{2\cos(3
s_1)}}\left(\cos(s_1)+\sqrt{3}\sin(s_1)\right)\,,\cr \cr
x_{\rm nn} & = & -{{3\cos(s_2)}\over{\cos(3 s_2)}}\,,\cr \cr
x_2 & = & {{3}\over{2\cos(3
s_2)}}\left(\cos(s_2)-\sqrt{3}\sin(s_2)\right)\,,\cr \cr
x_3 & = & {{3}\over{2\cos(3
s_2)}}\left(\cos(s_2)+\sqrt{3}\sin(s_2)\right)\,,
\end{eqnarray}
with $0\leq s_2\leq s_1\leq\pi/6$. It follows that
\begin{eqnarray}
\tilde V(r) & = & -{{16 \cos^2(3 s_1)\,\cos^2(3 s_2)}\over{729 x^2}}
(x-x_{\rm neg})(x-x_S)\times\cr \cr
& &\quad (x-x_H)
(x-x_{\rm nn})(x-x_2)(x-x_3)\,.
\end{eqnarray}
Take $r_i= 2M x_i$. The four positive roots are ordered as
follows: $r_S\leq r_2\leq r_3\leq r_H$. There are thus strings of the form of
our ansatz extending from $r_2$ to $r_3$, and outside the cosmological horizon
and inside the black hole horizon. The strings outside the cosmological horizon
are of infinite length, mass and action. The really relevant ones for our
purposes are those extending from $r_2$ to $r_3$, in complete analogy with the
results for Schwarzschild's black hole. We portray a numerical computation of
the classical Regge trajectory $W(m)$ in Fig. 2 for the case
$s_1=\pi/12$, that is, $H=\frac13 \sqrt{6}\, M $. Clearly to be seen are the
two branches
which had previously appeared for the rotating string in de Sitter spacetime.
Surprisingly enough, there is no minimum value for $W$ and $m$ greater than
zero in one of the branches, although it does appear in the second one. This
 is
due to the numerical integration, which is very inexact in the limit $T\to
T_{\rm min}(H)=6\pi M\sqrt{3}/\sin(3 s_1)$, and the fact is that there
{\it is} a minimum value for $W$, independent of $H$ and given by
$W_{\rm min}= 36\pi
M^2/\alpha'$, as can be found by computing the adequate limit $s_2\to0$;
the mass
$m$ also has a minimum value, but this time
$H$ dependent: $m_{\rm min}= 2\sqrt{3} M\sin(3 s_1) /\alpha'$. Notice that
 we
recover the results previously obtained for Schwarzschild spacetime.

\section{Conclusions}
We have seen that the study of the planetoid solutions to the classical
equations of motion of a string provides us with a variety of effects due to
the structure of the target spacetime. In particular there are two main effects
that we have uncovered:
\begin{enumerate}
\item the existence of a {\sl maximum} value for the angular
momentum of (equatorially) moving strings in spacetimes with particle horizons
(de Sitter and Schwarzschild-de Sitter in particular), which reflects itself
 on
the existence of two branches in  the  Regge plot. This means that
the number of bound  states  is finite in  the semiclassical
quantization. (But this finiteness must be exact, beyond the
semiclassical approximation).
\item the presence of a minimum value for the angular momentum in the case
 of a
black hole event horizon, as in Schwarzschild and Schwarzschild-de
 Sitter spacetimes.
\end{enumerate}
It is not difficult to understand this phenomenon in the light of elementary
quantum mechanics. In spacetimes with particle horizons it is necessary for
the
preservation of causality that if a string extends beyond the horizon that
 it be
infinite. The length is quantized in the same manner that the angular momentum
is, as can be read out from eq.(\ref{expresiones}); it is thus the
case that there
are a finite number of possible quantum planetoid strings.

As to the minimum value, given that if a string does penetrate into a
(Schwarzschild) event horizon and is to maintain its linearity it must extend
 to
infinity, we see that the ``cutting out" of part of the spacetimes is what
forces a minimum value even for classical values (quantum mechanically that
was only to be expected).

String solutions that generalize non-circular point particle
trajectories should also exist in the spacetimes considered
here. However, the $\sigma$ and $\tau$ dependence probably cannot
be separated as we did in the planetoid strings presented in this paper.

 We
want to stress that the Regge trajectories are no longer linear (even for weak
curvature) in the spacetimes considered here. We thus infer from this classical
test string calculations that the fundamental string spectrum will get strongly
modified in these non-trivial gravitational backgrounds.

\acknowledgments
ILE has to thank the LPTHE for their  hospitality on several occasions.

\newpage

\begin{center}

{\bf Figure Captions}

\end{center}

\bigskip

{\bf  Fig.1}: The reduced action $ W = 2\pi J $
(where $ J $ is the angular momentum) in units of $\pi M^2 / \alpha' $
 as a function of the string mass $ m $ in  Schwarzschild spacetime.

\bigskip

{\bf  Fig. 2}: The reduced action $ W = 2\pi J $
(where $ J $ is the angular
momentum) as a function of the string mass $ m $ in  Schwarzschild-de
Sitter spacetime.



 \end{document}